\documentclass[twocolumn,cleanfoot,cleanhead,10pt]{asme2e}

\usepackage{colortbl}
\usepackage[dvipsnames]{xcolor}
\usepackage{float}
\usepackage{graphicx}
\usepackage{array}
\usepackage{booktabs}
\usepackage{amsmath}
\usepackage{amsfonts}
\usepackage{amssymb}
\usepackage{multirow}
\usepackage{adjustbox}
\usepackage{tabularx}
\usepackage{cleveref}
\usepackage{subfigure}
\usepackage[dvipsnames]{xcolor}
\confshortname{ICONE 28}
\conffullname{ }

\confmonth{ }
\confdate{ }
\confyear{ }
\confcity{ }
\confcountry{ }

\papernum{ }

\title{Study on Laminar Turbulent Transition in Square Arrayed Rod Bundles}
\author{Carolina S. B. Dutra
    \affiliation{
	Ken and Mary Alice Lindquist\\ Department of Nuclear Engineering\\
	Pennsylvania State University\\
	University Park, PA 16802\\
    Email: cpd5488@psu.edu
    }	
}

\author{Elia Merzari
    \affiliation{
	Ken and Mary Alice Lindquist\\ Department of Nuclear Engineering\\
	Pennsylvania State University\\
	University Park, PA 16802\\
    Email: ebm5351@psu.edu
    }	
}

\begin{document}

\maketitle    

%%%%%%%%%%%%%%%%%%%%%%%%%%%%%%%%%%%%%%%%%%%%%%%%%%%%%%%%%%%%%%%%%%%%%%
\begin{abstract}
{\it The study of coolant flow behavior in rod bundles is of relevance to the design of nuclear reactors. Although laminar and turbulent flows have been researched extensively, there are still gaps in understanding the process of laminar-turbulent transition. Such a process may involve the formation of a gap vortex street as the consequence of a related linear instability. 
\par In the present work, a parametric study was performed to analyze the spatially developing turbulence in a simplified geometry setting. The geometry includes two square arrayed rod bundle subchannels with periodic boundary conditions in the cross-section.  The pitch-to-diameter ratios range from 1.05 to 1.20, and the length of the domain was selected to be 100 diameters.  No-slip condition at the wall, and inlet-outlet configuration were employed. Then, to investigate the stability of the flow, the Reynolds number was varied from 250 to 3000. The simulations were carried out using the spectral-element code Nek5000, with a Direct Numerical Simulation (DNS) approach. Data were analyzed to examine this Spatio-temporal developing instability. In particular, we evaluate the location of onset and spatial growth of the instability.

Keywords: instability, laminar-to-turbulent transition, DNS
}
\end{abstract}

%%%%%%%%%%%%%%%%%%%%%%%%%%%%%%%%%%%%%%%%%%%%%%%%%%%%%%%%%%%%%%%%%%%%%%
\begin{nomenclature}
\entry{$\rho$}{Density}
\entry{$D_H$}{Hydraulic diameter}
\entry{D}{Diameter}
\entry{DNS}{Direct Numerical Simulation}
\entry{E}{Turbulent Kinetic Energy}
\entry{$\mu$}{Viscosity}
\entry{p}{Pressure}
\entry{P}{Pitch}
\entry{Re}{Reynolds}
\entry{t}{Time}
\entry{u}{Velocity}
\entry{u'}{Fluctuation velocity}
\entry{z}{Streamwise direction}

\end{nomenclature}

%%%%%%%%%%%%%%%%%%%%%%%%%%%%%%%%%%%%%%%%%%%%%%%%%%%%%%%%%%%%%%%%%%%%%%
\section*{INTRODUCTION}

The study on laminar-turbulent transition dates back to the origins of fluid dynamics. Nonetheless, it remains an active area of research, as several open questions remain, especially for complex geometries.  

\par Even for pipe flows, which have been studied extensively through experiments \cite{Draad1995, Henningson1994, Sibulkin1962, Sreenivasan1982, Wygnanski1973} and more recently through numerical simulation \cite{Eckhardt2007, Moxey2010, Peixinho2013, He2015}, present unanswered questions. In fact, laminar pipe flow is unconditionally stable when examined through a linear stability analysis.  As such the critical Reynolds number or the Reynolds number above which laminar flow becomes unstable is subject to some controversy and multiple possible explanations have been provided. 

In this manuscript, we investigate in particular laminar-turbulent transition in rod bundles. The study of turbulence in this class of geometry is the importance of a variety of nuclear engineering applications, including reactor cores and heat exchangers. Meyer \cite{Meyer2010} provides an extensive review of the history of the understanding of turbulent flows for rod bundles.
A vortex street \cite{tavoularis2011rod} is generated in the narrow gaps between rods, driven by steep gradients of the streamwise velocity.
\par Several groups have investigated this gap vortex street experimentally in rod bundles and similar geometry \cite{choueiri2014experimental, guellouz2000structure, tavoularis2021further}. 
The current consensus indicated that these vortices are induced by a span-wise velocity gradient and that the vortex street increases the mixing. The average axial spacing of the vortices is independent of the Reynolds number and determined by the channel geometry. 
Recent studies have demonstrated that this vortex street is related to a linear instability that occurs at very low Reynolds numbers \cite{merzari2008biglobal, merzari2012optimal}. In eccentric annuli, a prototype for this class of geometry, the stability has been shown recently  to be present even at very low values of the eccentricity \cite{moradi2019flow, tavoularis2021further}. Key to this recent discovery was the observation that very long domains  may be required to trigger the instability at low eccentricity.
Furthermore, experimental, LES and DNS results have demonstrated that at low eccentricity the vortex street is not prevalent at sufficiently high Reynolds numbers \cite{merzari2009anisotropic, tavoularis2021further}. We also note that Taehwan et al. \cite{Taehwan2013} investigated experimentally the turbulent flow in square arrayed six-rod bundles and concluded that vortex trains only exists with small P/D.

Several questions therefore remain for laminar-turbulent transition in rod bundles: 
\begin{enumerate}
    \item Is the vortex street present at high P/D? 
    \item Does it play an important role during transition even if at higher Reynolds it is not dominant? If so, does it require a long domain to develop?
    \item What is the overall effect of P/D on transition and the critical Reynolds number?
\end{enumerate}

\par To attempt to address these issues we perform here a series of Direct Numerical Simulations (DNS) of turbulence. We examine the non-linear spatial-developing form of the instability problem. The setting is that of a numerical experiment: we assume an inlet/outlet configuration comprising two sub-channels. The inlet condition is assumed to be the fully-developed laminar distribution computer numerically. We then examine the spatial growth of the instability at different Reynolds numbers. 

\section*{METHODS}

\begin{figure}[h]
    \centering
    \includegraphics[scale=0.53]{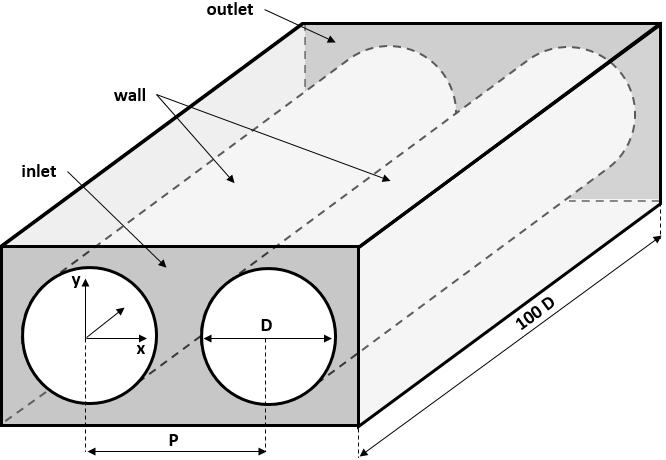}
    \caption{REPRESENTATION OF THE DOMAIN AND THE PHYSICAL MODEL}
    \label{figure_1} 
\end{figure}

\begin{figure}[h]
    \centering
    \includegraphics[scale=0.9552]{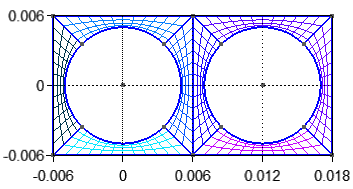}
    \caption{CROSS SECTION OF THE MESH OF P/D = 1.2}
    \label{figure_2} 
\end{figure}

The physical model consists of a two square arrayed rod bundle subchannel with pitch-to-diameter ratios ranging from 1.05 to 1.20. This three-dimensional model was built in the finite element mesh generator, Gmsh. There is a finer refinement near the wall to capture adequately the boundary layer effects. The diameter D of the fuel rod was kept constant while the pitch P was varied. The length of the channel $L$ was set to $L=100D$ and $L=200D$  to assure sufficient development of the instability - we will comment on this however as the length has important implications. An illustration of the geometric model and a cross-section representation of one of the hexahedral generated meshes are shown in Figures \ref{figure_1} and \ref{figure_2}, respectively.  

The Direct Numerical Simulations (DNS) were performed through the open-source Computational Fluid Dynamics (CFD) code, Nek5000, developed at Argonne National Laboratory \cite{nek5000-web-page}. This code uses the spectral-element method, a higher-order method ideally suited for DNS. Moreover, this method presents very little diffusion or dispersion, which is an important feature for stability calculations. 
Assuming the fluid is incompressible, the viscosity is constant, and there are no external forces, such as sinks, sources, and acceleration, the governing equations are given by Navier-Stokes Equations:
\par The Momentum Equation:
\begin{equation}
    \rho \left ( \frac{\partial \mathbf{u}}{\partial t} + \mathbf{u}\cdot \triangledown \mathbf{u}\right ) = -\triangledown p + \mu \Delta \mathbf{u}
\end{equation}

\par The Continuity Equation:
\begin{equation}
    \triangledown \cdot \mathbf{u} = 0
\end{equation}

The mesh generated for this case included $n=108,200$ elements ($L=100 D$) and $n=216,400$ ($L=200 D$), and the simulations were performed at the $N=5$ polynomial order. We verified that simulations performed using $N=3$ and $N=7$ yielded identical results to $N=5$.

For the simulations, the assumptions of the no-slip condition at the wall, periodic boundary conditions at the sides, inlet/outlet in the streamwise direction were made. For each P/D, a precursor laminar calculation was run and used to set the boundary condition at the inlet. The laminar profile was also used as the initial condition. These simulations had similar radial dimensions, but with the length of $1D$ in the streamwise direction. Periodic boundary conditions were used in the streamwise direction.  

The calculations were carried out for each of the geometries, with Reynolds numbers ranging from 250 to 3000, which is the main parameter to characterize the laminar-to-turbulent transition and is given by:
\begin{equation}
    Re = \frac{\rho u D_{H}}{\mu}
\end{equation}

\par The transition from the laminar to the turbulent regime happened naturally, without triggering the inflow with any disturbance. Yet, cases with small perturbances in the inlet were tested and did not present any relevant changes in the flow behavior and in the location of the instability.
\par Afterward, to quantify the behavior of unsteady pattern, the velocity fluctuations were obtained by subtracting the instantaneous velocity by the initial condition.
\begin{equation}
    u^{'}_i (x,y,z,t)=u_i(x,y,z,t)-<u_(x,y,z)>
    \label{eq4}
\end{equation}

From the velocity fluctuations, the turbulent kinetic energy was calculated at a given axial plane:

\begin{equation}
    E(z,t)=\frac{1}{\iint dxdy}\iint \frac{1}{2}\sum_{i=1,2,3}u^{'}_i (x,y,z,t)^{2}dxdy
    \label{eq5}
\end{equation}

We note that the simulations were conducted over very long transients over $t = 300 D/u$ convective units with a time step size of $dt = 10^{-3} D/u$. A second order time stepping scheme was used in time with $CFL=0.4$.

\section*{RESULTS}
\begin{figure}[h]
\centering
    \subfigure{\em(a) length of 100D}{
        \includegraphics[scale=0.51]{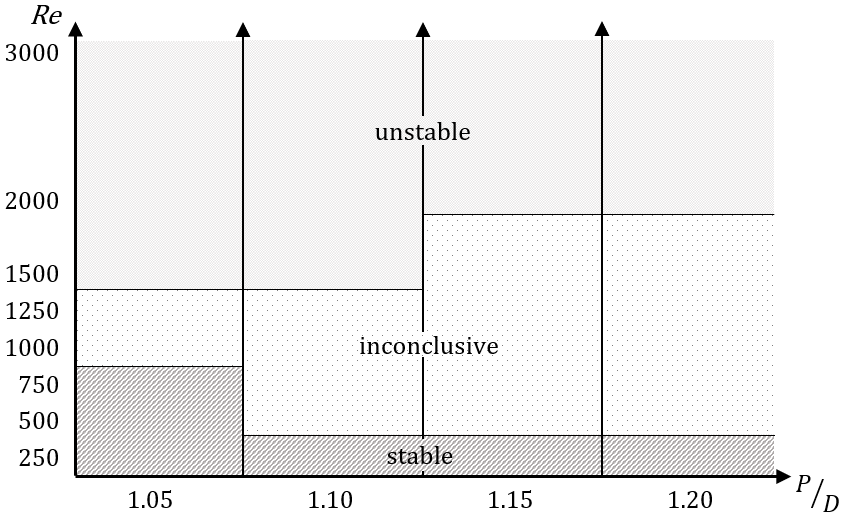}
        \label{figure3_a}
    }
\vskip 16pt
    \subfigure{\em(b) length of 200D}{
        \includegraphics[scale=0.51]{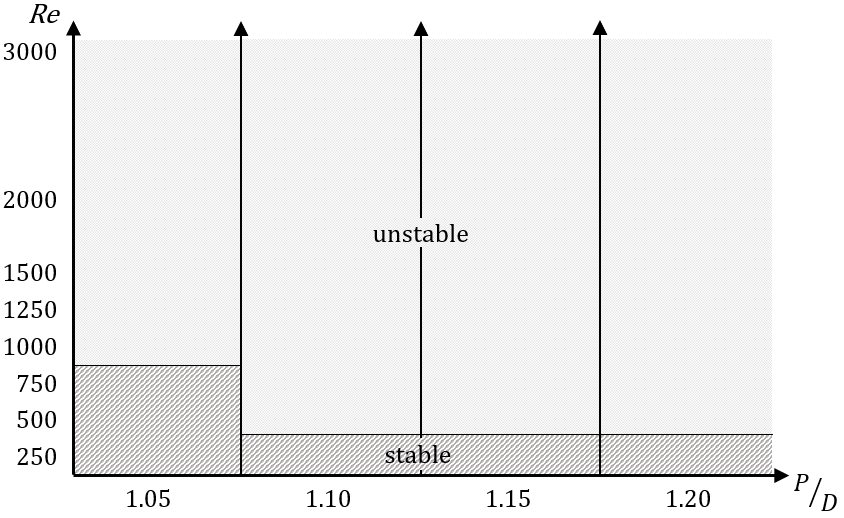}
        \label{figure_3b} 
    }
    \caption{DEPENDENCE OF STABILITY ON REYNOLDS NUMBER AND P/D}
    \label{figure_3}
\end{figure}

The stability results for the case with $L=100 D$ were classified into three categories: stable, inconclusive, and unstable (Figure~\ref{figure_3}). The stable state occurs at the lowest Reynolds numbers and is characterized by smooth flow paths. In other words, the laminar flow state remains unperturbed. In the unstable state, at time $t_{1}$, at some point downstream $z_{1}$, a strong oscillatory oscillation is observed, and a sustained unstable growth ensues for all time $t>t_{1}$ at all $z>z_{1}$.

However, we have also observed an inconclusive state. In this state, we observe an unstable growth at a given time $t_{1}$, which presents itself with characteristics similar to the unstable case. Still, this unstable growth disappears at later times, and after $t>t_{2}$, it decays toward the laminar profile. We note that the process could repeat itself, but given the long duration of these transients, we stopped the simulation at the end of the re-laminarizarion.

The interpretation of this re-laminarization, which occurs for all P/D examined, was not clear for the simulations performed with the channel length of $ 100D$. As pointed out by Moradi \cite{moradi2019flow}, the streamwise wavelength of the instability depends on the geometry and at high P/D  the streamwise length adopted may be insufficient. Furthermore, the growth rate of the principal mode of the instability may also be too weak to observe the vortex street.  In other words, this effect could be considered a numerical artifact related to the length of the domain. In fact,  as demonstrated by Lamarche-Gagnon and Tavoularis \cite{tavoularis2021further}, the channel length is crucial to observe the onset of turbulence at lower Reynolds numbers. This instability and subsequent re-laminarization could then be interpreted as a form of intermittency that occurs when the instability is not strong enough to fully transition at slightly supercritical conditions.

\par We then performed simulations with a domain twice as long for the lower range of Reynolds numbers to understand this inconclusive behavior. The results obtained by the simulations of both lengths, $100D$ and $200D$, respectively, were summarized in an instability map, as shown in Figure \ref{figure_3}. It was observed that the instabilities in the inconclusive range develop further downstream, therefore justifying the need for a longer channel.

\begin{figure}[h]
    \centering
    \includegraphics[scale=0.53]{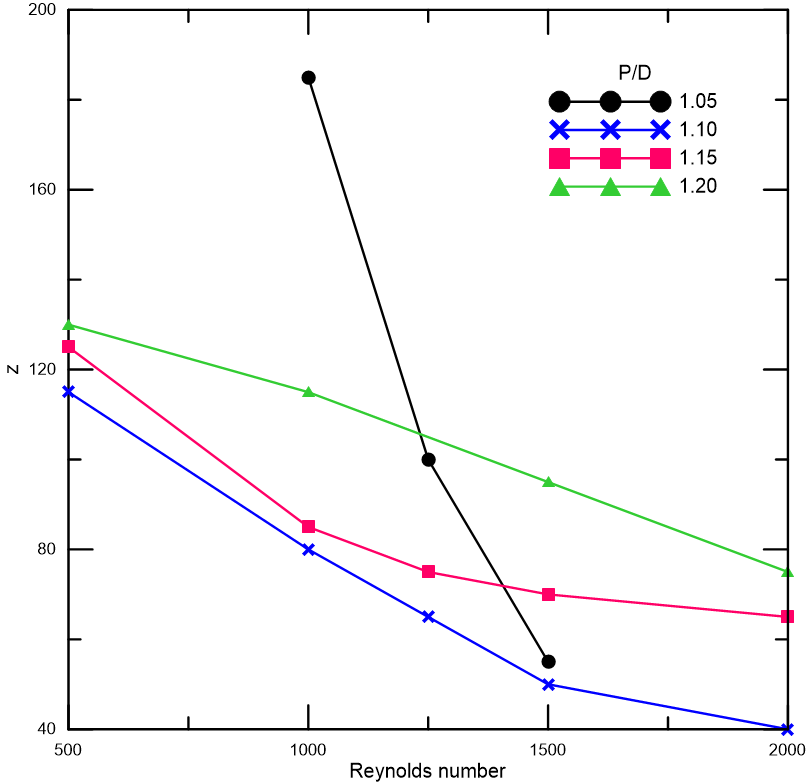}
    \caption{LENGTH AS FUNCTION OF THE CRITICAL REYNOLDS NUMBER AS AT THE LOCATION WHERE THE TRANSITION OCCURS FOR P/D = 1.05, 1.10, 1.15, AND 1.20}
    \label{figure_4} 
\end{figure}

Figure \ref{figure_4} shows the streamwise location of the spatial onset of the instability once the simulation reaches a steady-state, plotted as a function of the Reynolds number for each $P/D$. For the same Reynolds, the transition takes place in different positions for each $P/D$. Further, for every $P/D$, we note that the general trend points to a shorter transition as the Reynolds number increases. This is consistent with expectations as the growth rate of the instability will increase with the Reynolds number.  However, we also observe at the $P/D=1.05$ the trend is remarkably steeper than for higher P/D, reducing rapidly to much lower values than other P/D. This is consistent with the behavior in narrower gaps, where the small narrow gap may prevent the formation of the vortex due to enhanced viscous effects but, as the instability occurs, its intensity and growth rate increases (see again Lamarche-Gagnon and Tavoularis \cite{tavoularis2021further}).

The combination of these two observations (i.e., rapid change in onset and effect of length in predictions) may point us to why the vortex street has not been observed in rod bundles for high P/D. To do so would require undisturbed experimental lengths of the order of $200D$. At lower lengths, the transition occurs at higher Reynolds, where the vortex street for high $P/D$ is short-lived, as it transitions immediately to more chaotic behavior. This behavior will be discussed in more detail in the following.

\par We now examine in detail some of the flow fields computed and discuss qualitatively and quantitatively the process of transition observed. For the following analysis, as the secondary flow velocities represent less than 1\% of the axial velocity, and do not contribute significantly to the flow mixing, they were disregarded \cite{Meyer2010}. Besides, since high turbulent intensities occur in the narrow gap region, that was the region chosen to carry on with the analysis.

\par
\begin{figure}[h]
    \centering
    \includegraphics[scale=0.47]{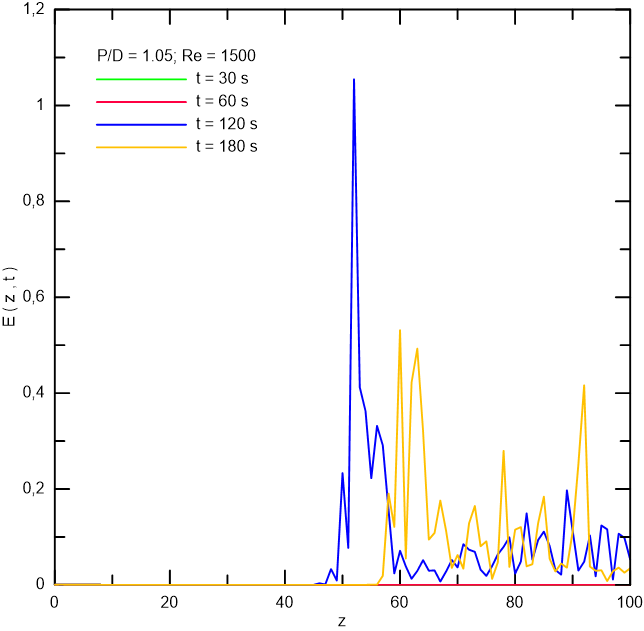}
    \caption{TURBULENT KINETIC ENERGY  AS A FUNCTION OF THE AXIAL DISTANCE FOR P/D=1.05 AND Re=1500}
    \label{figure_6} 
\end{figure}

\begin{figure}[h]
    \centering
    \includegraphics[scale=0.47]{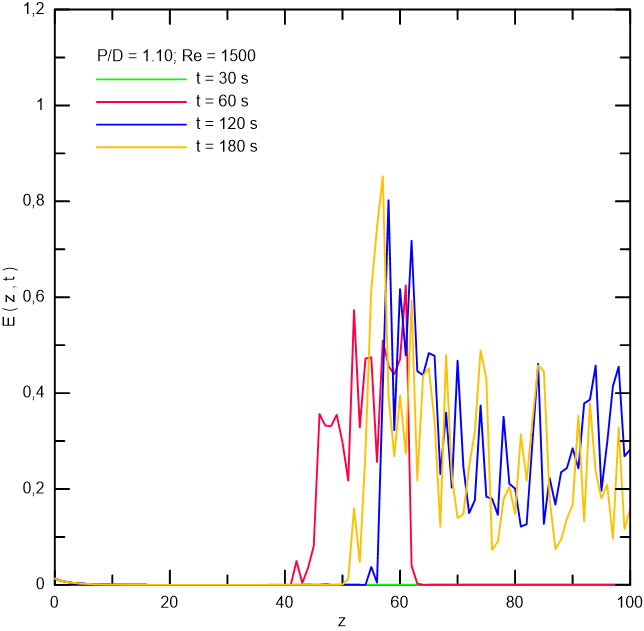}
    \caption{TURBULENT KINETIC ENERGY AS A FUNCTION OF THE AXIAL DISTANCE FOR P/D=1.10 AND Re=1500}
    \label{figure_7} 
\end{figure}

\begin{figure}[h]
    \centering
    \includegraphics[scale=0.47]{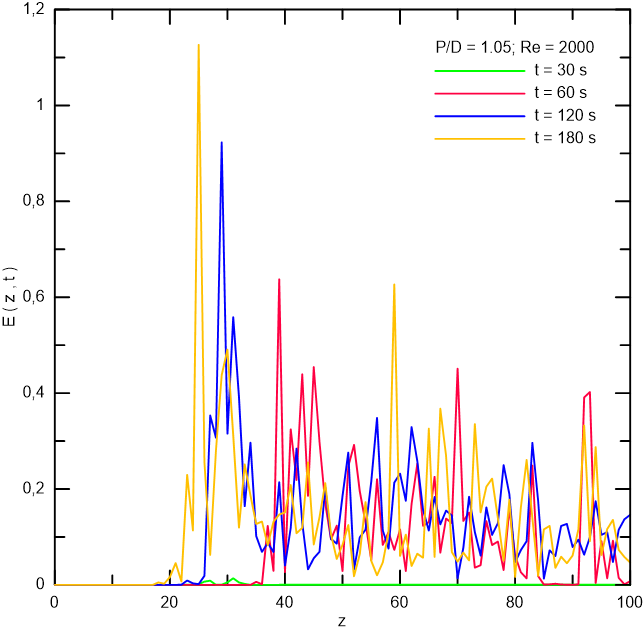}
    \caption{TURBULENT KINETIC ENERGY AS A FUNCTION OF THE AXIAL DISTANCE FOR P/D=1.05 AND Re=2000}
    \label{figure_8} 
\end{figure}

\begin{figure}[h]
    \centering
    \includegraphics[scale=0.47]{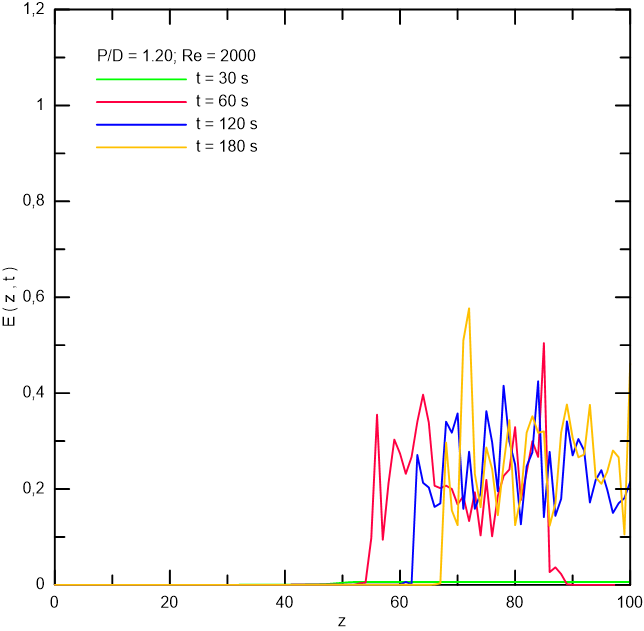}
    \caption{TURBULENT KINETIC ENERGY AS A FUNCTION OF THE AXIAL DISTANCE FOR P/D=1.20 AND Re=2000}
    \label{figure_9} 
\end{figure}

\begin{figure}[h]
    \centering
    \includegraphics[scale=0.47]{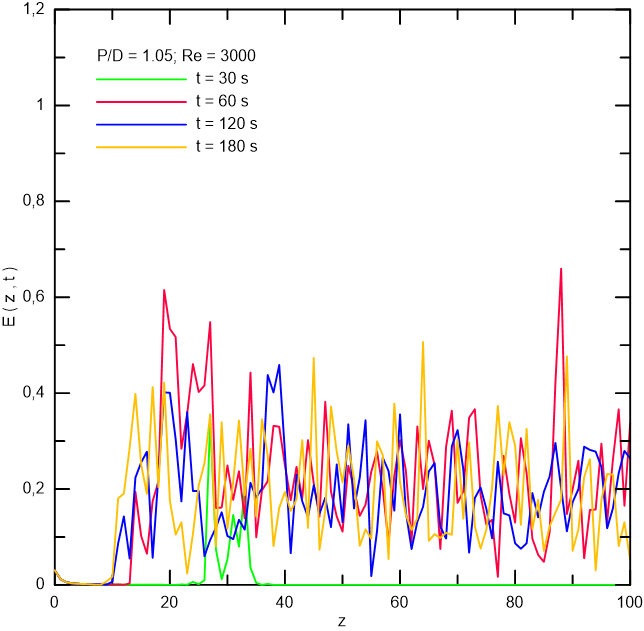}
    \caption{TURBULENT KINETIC ENERGY AS A FUNCTION OF THE AXIAL DISTANCE FOR P/D=1.05 AND Re=3000}
    \label{figure_10} 
\end{figure}

\begin{figure}[h]
    \centering
    \includegraphics[scale=0.47]{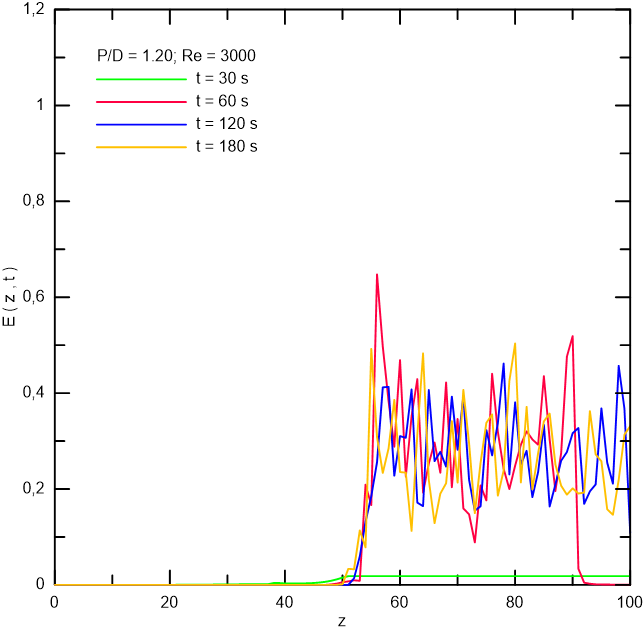}
    \caption{TURBULENT KINETIC ENERGY AS A FUNCTION OF THE AXIAL DISTANCE FOR P/D=1.20 AND Re=3000}
    \label{figure_11} 
\end{figure}

 We use the turbulent kinetic energy \eqref{eq5}  to study the onset of turbulence. The turbulent kinetic energy as a function of the axial distance for the times of $30$, $60$, $120$, and $180 D/u$ convective units, are shown in Figures \ref{figure_6} to \ref{figure_11}. 
\par It can be observed the existence of high velocity jumps between laminar and turbulent profiles and, therefore, large fluctuations, yielding great peaks of energy. For the same Reynolds number, the energy peaks were higher for $P/D=1.05$ rather than $P/D = 1.2$, allowing us to conclude that the decrease of the gap space significantly enhances the turbulent kinetic energy. Also, the oscillations start later and become less dominant at higher P/D.  
\par
\begin{figure*}
\centering
    \subfigure{\em(a) P/D = 1.1, y = P/2}{
        \includegraphics[scale=0.965]{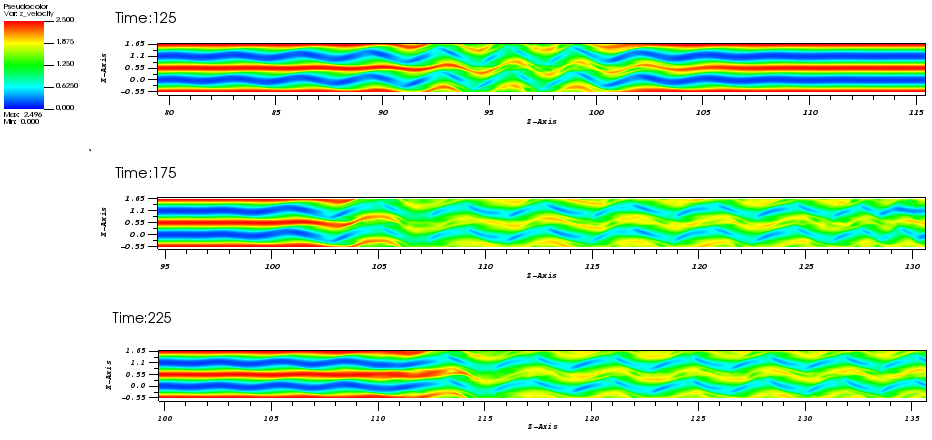}
        \label{figure12_a}
    }
\vskip 16pt
    \subfigure{\em(b) P/D = 1.2, y = P/2}{
        \includegraphics[scale=0.965]{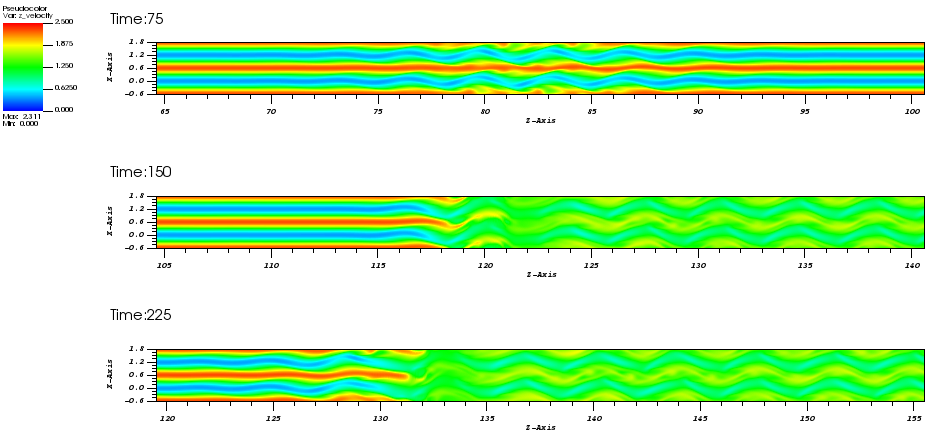}
        \label{figure_12b} 
    }
    \caption{STREAMWISE VELOCITY IN CUTS OF THE PLANE $y=P/2$ FOR Re = 500}
    \label{figure_12}
\end{figure*}

 Furthermore, screenshots of the streamwise velocity were taken for Re = 500 at the plane $y=P/2$, for $P/D = 1.10$ and $P/D = 1.20$, respectively, as shown in Figure \ref{figure_12}.


\par The axial fluctuating velocity increases as localized unstable spots are generated, grow, and then are merged with each other occupying the domain. This process determines the beginning of the transitional state. At some point in time, the intermittency becomes less prominent and the motions more uniform. This time lag between the beginning of the simulation and the formation of a spatial instability starting from a fixed streamwise location depends on the $P/D$ and the Reynolds number. The axial starting point of the instability is defined as the location where the energy increases rapidly to its steady value, as observed in Figure~\ref{figure_11}. Our primary interest is the evaluation of the spatial instability after this initial transient. 

\par
\begin{figure*}
\centering
    \includegraphics[scale=0.95]{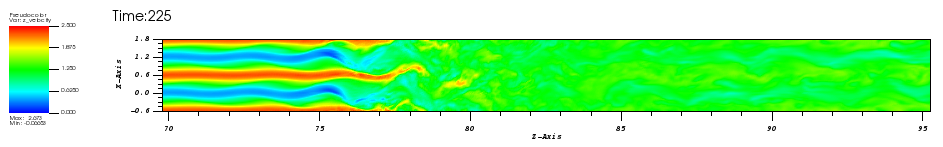}
    \caption{STREAMWISE VELOCITY IN THE PLANE $y=P/2$ FOR P/D = 1.2 AND Re = 2000}
    \label{figure_13}
\end{figure*}

 The formation of a gap vortex street is evident, in particular at low $P/D$, with the typical sinusoidal pattern for the streamwise velocity. We note that the gap vortex street is present at all $P/D$, but at the higher $P/D$, it is far less dominant. In fact, while they are always present, they at high $P/D$ and high $Re$ they quickly give way to a chaotic state, and the sinusoidal pattern is not recognizable soon after onset (e.g., Figure~\ref{figure_13}). 

\par It is possible to notice that the transition occurs typically earlier spatially for the lower $P/D$ cases, although the laminar solution was stable at higher Reynolds numbers. Furthermore, the maximum cross velocity, and therefore the mixing, were higher due to the presence of strong vortex trains. As the Reynolds number increases, the growth rate increases, and the time to reach a turbulent state was reduced for all $P/D$.

%%%%%%%%%%%%%%%%%%%%%%%%%%%%%%%%%%%%%%%%%%%%%%%%%%%%%%%%%%%%%%%%%%%%%%
\section*{CONCLUSION}
\par In this work, the onset of turbulence in a rod bundle case was modeled using Direct Numerical Simulation. We focused on the spatial growth of the instability in the streamwise direction given a fully-developed laminar inlet. We confirmed that the Reynolds number and the channel geometry are essential parameters for the transient characterization. Above a critical Reynolds number, we observed the presence of a gap instability.  This study demonstrates for the first time, to our knowledge, that this type of instability can occur even at fairly large $P/D$ and very low Reynolds provided the domain is sufficiently long. 
\par For shorter domains (e.g., $L=100D$),  the transition occurred at a lower axial distance, for smaller $P/D$, the oscillations were dominant, and strong vortex trains were observed.  For higher $P/D$, the vortex street was also present, but occurred at higher Reynolds and its dominance is clearly less prominent than for lower P/D. This latter item is consistent with previous observations \cite{merzari2011proper}, and the mechanism for this rapid collapse in energy content of the vortex street will be the subject of a future study.
\par For longer lengths ($L=200D$), which can accommodate very slow spatial growth of unstable modes, the vortex street also occurred at higher P/D, and even lower Reynolds. This is consistent with very recent observations by Lamarche-Gagnon and Tavoularis \cite{tavoularis2021further} for eccentric annuli.  At higher Reynolds and high $P/D$ the vortex street becomes rapidly less dominant with a behavior similar to what was observed at lower lengths. 
\par In conclusion, the present work point to the gap instability as the primary form of laminar-turbulent transition for rod bundles regardless of the $P/D$. The length of the bundle,  numerical or experimental, can affect significantly whether the vortex street can be observed in practice. 
%%%%%%%%%%%%%%%%%%%%%%%%%%%%%%%%%%%%%%%%%%%%%%%%%%%%%%%%%%%%%%%%%%%%%%

\bibliographystyle{asmems4}

%%%%%%%%%%%%%%%%%%%%%%%%%%%%%%%%%%%%%%%%%%%%%%%%%%%%%%%%%%%%%%%%%%%%%%
% The bibliography is stored in an external database file
% in the BibTeX format (file_name.bib).  The bibliography is
% created by the following command and it will appear in this
% position in the document. You may, of course, create your
% own bibliography by using thebibliography environment as in
%
% \begin{thebibliography}{12}
% ...
% \bibitem{itemreference} D. E. Knudsen.
% {\em 1966 World Bnus Almanac.}
% {Permafrost Press, Novosibirsk.}
% ...
% \end{thebibliography}

% Here's where you specify the bibliography database file.
% The full file name of the bibliography database for this
% article is asme2e.bib. The name for your database is up
% to you.
\bibliography{asme2e}

\end{document}